# The Signatures of Voids and the CMBR

Sharon L. Vadas,[a]

*Center for Particle Astrophysics, 301 Le Conte Hall*

*University of California; Berkeley, CA 94720-7304*

We explore the signature of a cold dark matter compensated void in the quasi non-linear regime. We find that this void is entirely a cold spot on the microwave background, in contrast to a non-linear void. On the last scattering surface(LSS), it appears as either a hot or cold spot depending on where this surface cuts the void. In addition, because the usual cancellations do not occur, the void's LSS signature can be very large as it is proportional to $R/H^{-1}$ rather than $(R/H^{-1})^3$. This implies strict limits for voids on the LSS.

## I. Introduction

A void is a region with underdensity $\delta \equiv 1 - \rho_{in}/\rho_{out}$ and radius $R$, where $\rho_{in}$ and $\rho_{out}$ are the energy densities inside and outside the void, respectively. A cold dark matter (CDM) void has three evolutionary phases: linear($\delta \ll 1$), quasi non-linear($\delta \simeq .1$ to .9) and non-linear($\delta \simeq 1$). In the quasi non-linear regime the void deepens quickly and its wall thins. A void's signature (e.g. the Rees-Sciama effect[1]) is obtained by tracing geodesics as the void evolves; for a compensated, non-linear void[2], it is

$$\Delta T/T = .4(c^{-1}R_e/H_e^{-1})^3[1 - (5/3)\cos^2\psi_0]\cos\psi_0, \tag{1.1}$$

where $H$ is Hubble's constant and "e" is when the photon exits the void. Also, $\psi_0$ is the angle between the photon's direction and a line from the void's center to the photon at $t_e$. Note that a non-linear void is a cold spot ($\psi_0 < 39°$) surrounded by a hot ring.

## II. Fluid and Geodesic Equations

The spherically symmetric, comoving metric we use here is

$$ds^2 = -c^2\Phi^2(t,r)dt^2 + \Lambda^2(t,r)dr^2 + R^2(t,r)(d\theta^2 + \sin^2\theta d\phi^2). \tag{2.1}$$

We refer the reader to Ref. [3] for the hydrodynamic equations and an explanation of how the code works. We define the comoving-frame radius $R_{\rm CF} = R(t_i, r)$, so that at $t$, the sphere labeled by $r$ has physical radius $R(t)$ and comoving-frame radius $R_{\rm CF}(t)$. In addition, $Z_{\rm CF} = R_{\rm CF}\sin\theta$ and $X_{\rm CF} = R_{\rm CF}\cos\theta\cos\phi$.

The geodesic equations for a photon can be shown to be[4], $z = dr/dt$, $w = d\theta/dt$,

$$\frac{dw}{dt} = w\left[-\frac{2}{R}(\Phi U + R'z) + \frac{\dot\Phi}{\Phi} + \frac{R'U'}{\Gamma^2\Phi}z^2 + \frac{2\Phi'}{\Phi}z + \frac{RU}{\Phi}w^2\right], \tag{2.2}$$

$$\frac{dz}{dt} = z\left(\frac{\dot\Phi}{\Phi} + \frac{R'U'}{\Gamma^2\Phi}z^2 + \frac{2\Phi'}{\Phi}z + \frac{RU}{\Phi}w^2\right) - \frac{2\Phi U'}{R'}z - \frac{\Phi\Phi'\Gamma^2}{(R')^2} - \left(\frac{R''}{R'} - \frac{\Gamma'}{\Gamma}\right)z^2$$

$$+ \Gamma^2 R\, w^2/R', \tag{2.3}$$

where $U = \dot R/\Phi$ and $\Gamma = R'/\Lambda$. In addition, the energy of a photon as measured by comoving observers at each photon's location is

$$E(t) = E(t_i)\frac{\Phi(t)}{\Phi(t_i)}\frac{w(t_i)}{w(t)}\left(\frac{R(t_i)}{R(t)}\right)^2, \tag{2.4}$$

where $E(t_i)$ is the photon's initial energy and the propagation is non-radial ($w \neq 0$).

---

[a] vasha@physics.berkeley.edu; Contributed talk: 17[th] Texas Symposium for Relativistic Astrophysics

### III. Signature of a Void

Consider a compensated, quasi non-linear void with initially unperturbed velocity in a flat, Friedemann-Robertson-Walker (FRW) universe. The void and void wall expand outward faster than the expansion rate of the universe. There are three major effects contributing to the temperature distortion of a photon as it enters, crosses and exits the void. As it enters and exits, the distortion is blueshifted because the photon jumps into frames moving towards it. And as it crosses, it is redshifted because the photon moves into frames expanding away from it faster than the expansion rate of the universe.

Suppose a photon with energy $E(t)$ moves through any part of this void region from $t_1$ to $t_2$. At the same time, another photon travels outside the void region. The temperature distortion of the former photon at time $t_2$ is

$$\frac{\Delta T}{T}(t_1, t_2) = \frac{E(t_2)}{E(t_1)} \frac{a(t_2)}{a(t_1)} - 1, \qquad (3.1)$$

where $a(t)$ is the scale factor of the universe. (These two photons do not move the same distance in general). We apply this to the inner region of a void which expands outward as $t^\alpha$ with $\alpha = 2/3 + 2\delta/9$ for $\delta < 1$ [4]. Let $t_i$ and $t_f$ be the initial and final times, and let $t_1$ and $t_2$ be the times the photon leaves the wall and enters the void, and leaves the void and enters the wall, respectively. The total temperature distortion is

$$\frac{\Delta T}{T}(t_i, t_f) \simeq \frac{\Delta T}{T}(t_i, t_1) - \frac{2}{3}\delta(t_2)\frac{R(t_2)}{cH(t_2)^{-1}} + \frac{\Delta T}{T}(t_2, t_f). \qquad (3.2)$$

The first (third) term is the contribution from the first (second) void wall. The second term shows that the redshifting acquired after crossing the void is first-order in $R/H^{-1}$, and depends only on $\delta$, not on details of the void wall. It also increases linearly with distance across the void. Because the sum of the terms in Eq. (3.2) is third order for a non-linear void (see Eq. (1.1)), the temperature distortions acquired entering and leaving the void must also be first-order. We verify these results numerically.

Place a void with radius $R_{\text{void}}(t_i)$ at the origin, and an observer far outside the void on the $+Z_{\text{CF}}$ axis. Initially all photons have $\phi = 0$, start at the same value of $Z_{\text{CF}}$, and propagate parallel to the $Z_{\text{CF}}$-axis. They always remain in the $X$-$Z$ plane, by symmetry.

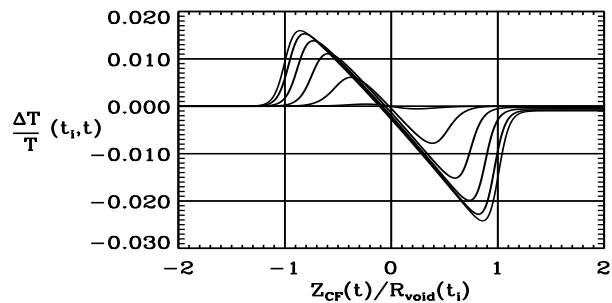

Figure 1: Temperature distortions of photons crossing an evolving void.

In Figure 1, we plot the temperature distortion, $\Delta T(t_i, t)/T$, as a function of $Z_{\text{CF}}(t)/R_{\text{void}}(t_i)$ for photons which pass through the void with $X_{\text{CF}}(t_i)/R_{\text{void}}(t_i) = .1, .3, .5, ..., 1.1$.

Table 1: Compensated Void with $\delta_e \simeq .4$ and $c^{-1}R_e/H_e^{-1} \simeq .2$

| $X_{\mathrm{CF}}(t_i)/R_{\mathrm{void}}(t_i)$ | $(c^{-1}R_e/H_e^{-1})^{-3}\Delta T(t_i, t_f)/T$ | |
|---|---|---|
| | Quasi non-linear | Non-linear |
| .1 | $-(.106 \pm .001)$ | $-.259$ |
| .3 | $-(.087 \pm .001)$ | $-.197$ |
| .5 | $-(.057 \pm .001)$ | $-.0866$ |
| .7 | $-(.0226 \pm .0005)$ | $.0428$ |
| .9 | $(.216 \pm .005)E - 2$ | $.119$ |
| 1.1 | $-(.23 \pm .01)E - 3$ | $0.0$ |

(See Ref. [4] for details). Upon exiting the void, $c^{-1}R_e/H_e^{-1} \simeq .21$ and $\delta_e = .44$. (The exit location is where $\rho/\rho_{out}$ is a maximum in the second void wall). As predicted, a photon's energy is blueshifted upon entering and leaving the void, and is redshifted while crossing the void. The curve with the largest temperature distortions has $X_{\mathrm{CF}}(t_i)/R_{\mathrm{void}}(t_i) = .1$. Using Eq. (3.2), $\Delta T(t_1, t_2)/T \simeq .06$, similar to Figure 1's more complicated "value" of .04. As $X_{\mathrm{CF}}(t_i)$ increases, the temperature distortions decrease because the parallel component of the wall's velocity decreases and the distance across the void decreases.

The final temperature distortions are much smaller than those obtained en-route (see Figure 1), and are listed in Table I along with the non-linear results from Eq. (1.1). The distortion is largest near the void's center, and is smaller than in the non-linear case because $\delta_e$ is not large enough[4]. In addition, the entire void appears as a cold spot. The large, hot ring is missing because the wall is thick, preventing the blueshifting from dominating over the the redshifting that occurs while crossing the fuzzy edge of the void.

### IV. Signatures on the Last Scattering Surface

The optical depth of the LSS is $\tau(t_1, t_0) = \int_{t_1}^{t_0} \sigma_T n_{elec} c \, dt$, where $n_{elec}$ is the electron density, $\sigma_T$ is Thompson's cross section, and $t_0$ is the observer's time. We assume that the LSS is instantaneous; once $\tau(t_1, t_0)$ drops below one, the number of interactions between photons and electrons is negligible. Then the LSS has an optical depth of one.

Define $t_f$ to be the time all LSS photons completely leave the void region. The optical depth between $t_1$ and $t_f$ is $\tau(t_1, t_f) = cg\sigma_T \int_{t_1}^{t_f} n(t, X_{\mathrm{CF}}, Z_{\mathrm{CF}}) dt$, where we assume that the light traces the mass: $n_{elec} = gn$. But because $\tau(t_1, t_f) + \tau(t_f, t_0) = 1$, we can move the void relative to the LSS (and therefore sample any part of the void) by changing $\tau(t_f, t_0)$ (i.e. moving the observer toward or away from the void). The last scattering surfaces then, are surfaces of constant optical depth, $\tau(t_1, t_f)$.

We can deduce the spatial geometry of the LSS when it crosses a void. Because the void is underdense, photons must travel farther inside than outside the void in order to have the same optical depth. Thus, photons that last scatter in the void are emitted earlier in time and therefore travel farther. If we plot the LSS in comoving-frame coordinates, the LSS curves away from the observer in the direction of the void.

To a given LSS, we apply the following transformation. Using Eq. (3.1) with $t_2$ replaced by $t_f$, the temperature distortion of a photon which last scattered at time $t_1$ is

$$\frac{\Delta T}{T}(t_1, t_f) = \frac{E(t_f)a(t_f)}{E(t_i)a(t_i)} \frac{E(t_i)a(t_i)}{E(t_1)a(t_1)} - 1 = \frac{\frac{\Delta T}{T}(t_i, t_f) - \frac{\Delta T}{T}(t_i, t_1)}{1 + \frac{\Delta T}{T}(t_i, t_1)}. \quad (4.1)$$

For most LSS surfaces, $|\Delta T(t_i,t_f)/T| \ll |\Delta T(t_i,t_1)/T| \ll 1$ so that $\Delta T(t_1,t_f)/T \simeq -\Delta T(t_i,t_1)/T$; whatever temperature distortion is acquired getting into the void region at $t_1$ is permanently frozen into the LSS with an accompanying negative sign. In general then, the first-order effects shown in Figure 1 are frozen into the LSS, resulting in relatively large signatures. In Figure 2, we show the signature of the LSS, $\Delta T(t_1,t_f)/T$, as a function of $Z_{CF}{}^{out}/R_{void}(t_i)$ and $X_{CF}(t_i)/R_{void}(t_i)$, where $Z_{CF}{}^{out}$ is the value of $Z_{CF}(t)$ for a photon outside the void region on the same last scattering surface (i.e. constant optical depth). (A LSS can be represented uniquely by its value of $Z_{CF}{}^{out}$). It is clear that the sign and magnitude of the signature depends sensitively on where the LSS slice the void. If the void is sliced near the back (front), the void is a cold (hot) spot on the microwave background. At the peaks of the distribution, the enhancement over $\Delta T(t_i,t_f)/T$ is $25 \simeq 2.5\delta_e(c^{-1}R_e/H_e^{-1})^2$. Only in one area is the signature relatively small.

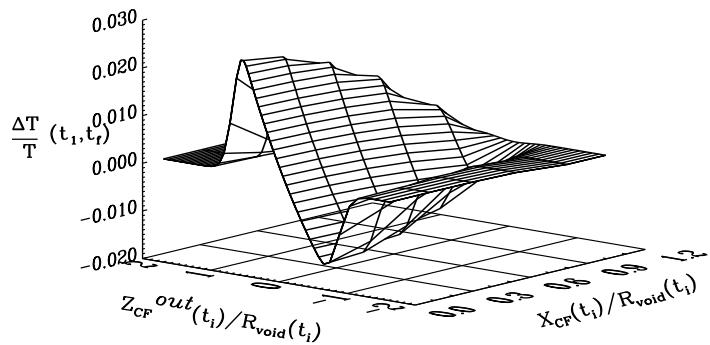

Figure 2: Signature of a void on the LSS—depends on where the LSS slices the void.

### V. Discussion

In conclusion, when a quasi non-linear void lies in front of the LSS, it appears entirely as a cold spot on the microwave background. However, when this same void lies on the LSS, its signature depends on where the LSS cuts the void; the void can appear to be hot or cold. In addition, the void's LSS signature is much larger than its signature in front of the LSS because it is first-order in $R/H^{-1}$. Over the scales that $\Delta T/T \lesssim 3 \times 10^{-5}$ experimentally, quasi non-linear voids with $\delta \simeq .3$ can be roughly limited to $c^{-1}R/H^{-1} \lesssim 10^{-4}$; they must be very small.

---

S.L.Vadas was supported by the President's Postdoctoral Fellowship Program at the University of California, and NSF Grant AST-9120005.